# Towards Holistic Prompt Craft

**Joseph Lindley**
Imagination Lancaster
Lancaster University, UK
j.lindley@lancaster.ac.uk

**Roger Whitham**
Imagination Lancaster
Lancaster University, UK
r.whitham@lancaster.ac.uk

FIG 1.1

## ABSTRACT

We present an account of an ongoing practice-based Design Research programme that explores the interaction affordances of real-time AI image generators. Based on our experiences from three installations, we reflect on the design of PromptJ, a user interface built around the concept of a prompt mixer. Our first contribution is a series of strong concepts based on our reflections of designing and deploying PromptJ. We cohere and abstract our strong concepts into the notion of *Holistic Prompt Craf*t, which describes the importance of considering all relevant parameters concurrently. Finally, we present PromptTank, a prototype design which exemplifies the principles of Holistic Prompt Craft. Our contributions are articulated as strong concepts or intermediate knowledge that are intended to inform and inspire practitioners and researchers who are designing with image generation models or developing novel interaction paradigms for generative AI systems more generally.

### CSS Concepts

• Human-centered computing~Interaction design

## INTRODUCTION

We are in the midst of a paradigm shift in how computers are used, which is being driven, in part, by the ubiquitous adoption of Generative AI (GenAI). This work contributes to this paradigm shift, focusing specifically on exploring the affordances of image-oriented diffusion-based GenAI systems (the examples in this paper are all built with Stable Diffusion). To do this, we provide a reflexive account of the development, use, and adaptation of a prototype user interface called PromptJ. PromptJ was created as the interface for an interactive installation, Shadowplay [1]. Shadowplay leverages an image-to-image workflow to allow users, participants or visitors to use their bodies to create AI imagery in real-time by blending a visual input from a camera with text prompts. There are a wide range of images showing this throughout the pictorial (e.g., Fig 1.1 above).

Methodologically speaking, this work is an example of practice-based Design Research (synonymous with a Research through Design approach). Practice-based Design Research is particularly well-suited to exploring the dynamic and rapidly shifting context of GenAI adoption [3]. Beyond Design Research's flexibility and reflexivity, a growing research programme demonstrates how the challenges of producing insights relating to GenAI's adoption align well with key Design Research strategies (such as those described in [6]). Examples of these strategies in practice can be seen in contemporary research into AI around themes of uncertainty [2], planetary AI [4], the notion of AI prompt craft [1], and the materiality of AI [5, 7]. The contributions we offer through this reflexive account of the practice-based Design Research process should be considered as "strong concepts" [8], i.e., they are insights derived from specific instances but whose utility is as generative or inspirational building blocks for other practitioners and researchers working in overlapping or related areas.

We offer two contributions. First, we describe several strong concepts relating to user interface (UI) design for interacting with GenAI based on our in the wild experiments with the PromptJ UI. These insights are based on our experiences in the design and use of the PromptJ UI concept. While this means our strong concepts are rooted in our experience working with a specific technology stack, there is direct relevance to the majority of diffusion-based GenAI systems and arguably there is also relevance to the vast array of other GenAI systems. We cohere and combine our strong concepts into the Holistic Prompt Craft Framing. Our second contribution is the presentation of PromptTank, a reimagining of an image generator UI which unifies the previously introduced strong concepts under the framing of Holistic Prompt Craft in an experimental spatial design metaphor.

## Shadowplay

At the centre of this work is the Shadowplay project, which is an interactive installation, performance system, and novel means to interact with image generation [1, 1]. In response to the rapid innovation in GenAI and also the wide variety of contexts which we have exhibited and demonstrated Shadowplay, the UI at the centre of the project is rapidly evolving. This pictorial focuses on the current version of that UI, which is named PromptJ.

### Light prompting set up

Since mid 2024 we have used a configuration of Shadowplay (Fig 2.1, 2.2) that captures a shadow from a rear-projection screen using a camera, processes this, then projects the output on an identical adjacent screen. There are some variations across our installations, but they mostly follow the configuration shown in Fig 2.2.

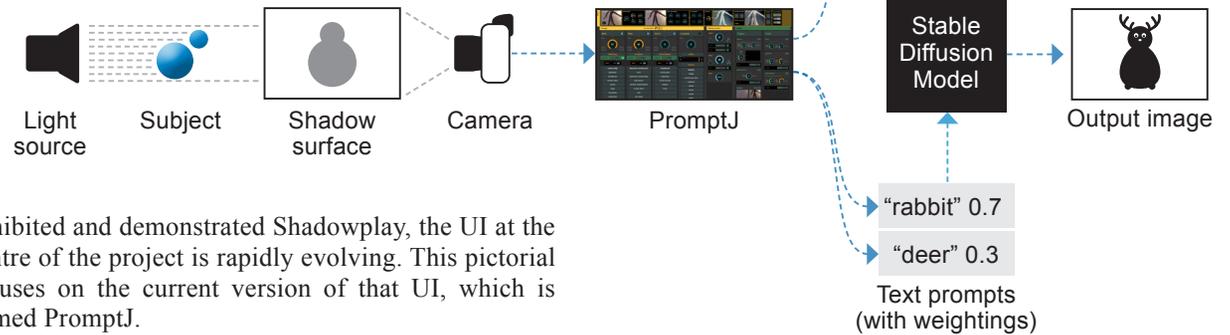

FIG 2.1

### PromptJ

PromptJ built upon an earlier version of the interface which leveraged pre-scripted scenes [1]. These scenes would use predefined sets of text prompts, with timed transitions between them, to create a positive experience with Shadowplay. The concept of PromptJ was to adopt the skeuomorph of an audio mixer. In doing so we wished to explore the potential to playfully improvise with how text prompts are used in Shadowplay, in a way that is both visually (in terms of UI design) and experientially reminiscent of the way a DJ or VJ manipulates sound and visuals in live performance contexts. PromptJ and its accompanying backend is built in Touch Designer (see https://derivative.ca/), utilises StreamDiffusion for real-time image generation (see https://github.com/cumulo-autumn/StreamDiffusion), and uses the SDXL Turbo model (see https://huggingface.co/stabilityai/sdxl-turbo).

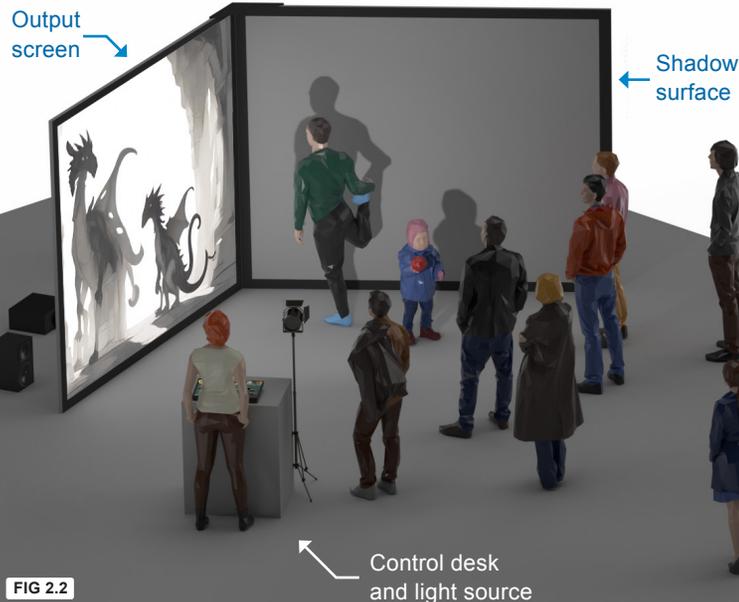

FIG 2.2

**THE PROMPTJ UI**

*Image prompt controls*
The upper strip of the UI is a horizontal chain that shows the signal path from the raw input image, to a post-processed image, to eventually the AI-generated output.

*Text prompt controls*
The lower part of the UI provides control over text-based prompting, some model settings and a series of triggers and timers that can be used for automation.

We wanted to create an user experience which shared in the interactional and visual languages of other live performance tools, with circular encoders, left-to-right signal flows and vertical playlists of content.

*Noise controls*
Controls for a static noise layered that can be overlayed on the camera input.

*Image adjustment controls*
Controls for generic brightness, contrast and gamma adjustments, colourisation and a crossfader to bring in and out a predefined image processing network (Salford Mode)

*Output preview and other controls*
The image chain ends with controls for stopping and starting the model and controlling automated capture of the output images as snapshots.

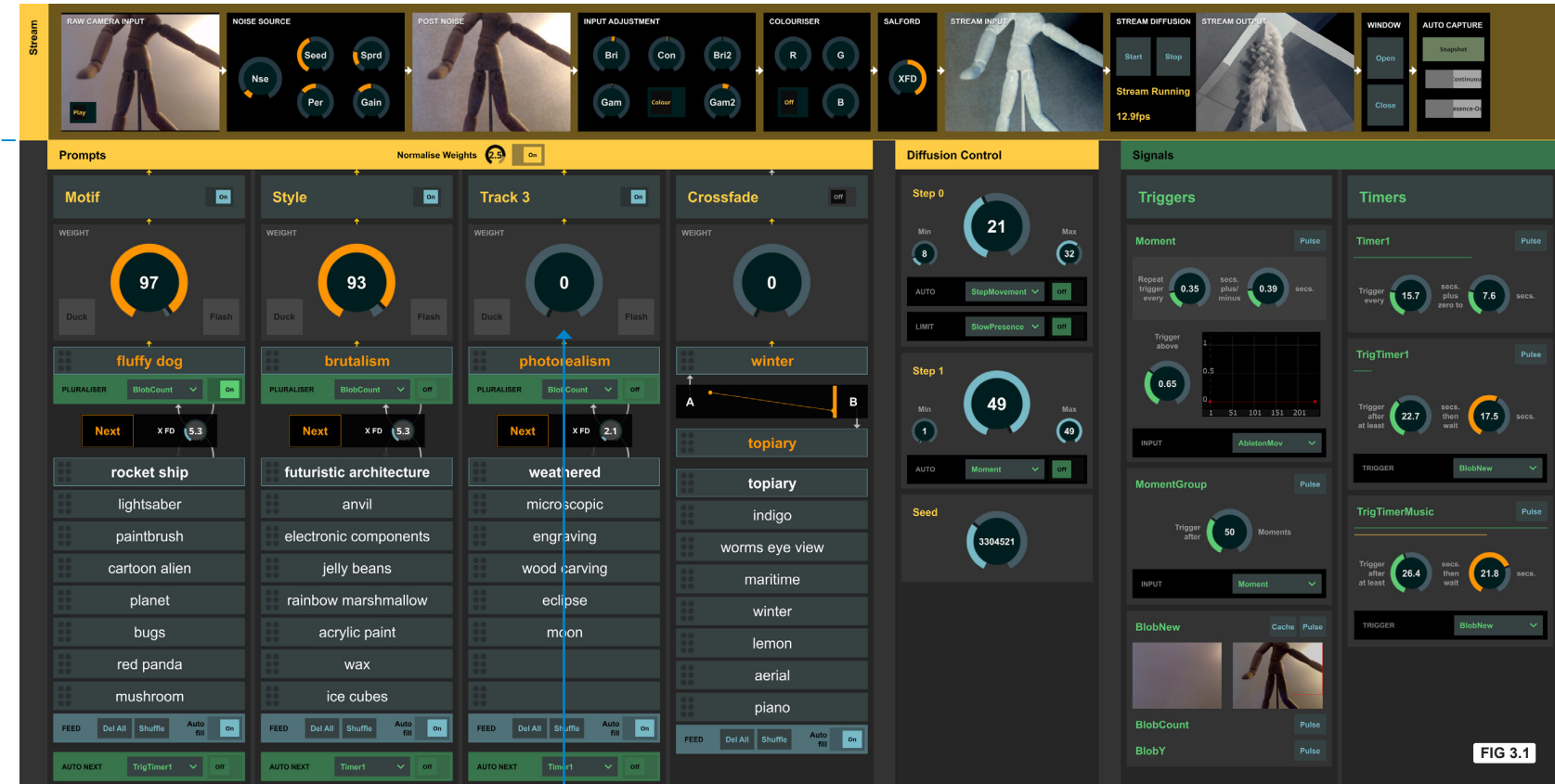

*Text prompt tracks*
This part of the UI mimics an audio mixer, with a series of separate tracks, each its with own textual content and weighting (shown by the fader above). The active prompt appears in orange, and below this is a playlist of prompts that can be selected from to stepped through in series.

*Weight controls*
Priority is given to the text prompt weighting controls, as during prompt crafting, multiple prompts need to be balanced against one another.

*Model controls*
The UI surfaces controls for the diffusion level (i.e., strength of the AI) and seed.

*Triggers and Timers*
Controls for signals that can be used to put PromptJ into an 'autopilot' mode, with event and time-based parameter changes.

FIG 3.1

**Page Numbers will be added here and either centered or right-aligned**

**PROMPTJ DEPLOYMENTS**

In this pictorial we refer to three specific installations of Shadowplay. All of these used the PromptJ interface to control our bespoke image-to-image generation system and took place in different UK-based locations (Fig 4.4). The strong concepts presented in this paper are derived from reflections on designing PromptJ, then using and adapting it across these three distinct contexts.

**[blank for review] Festival**

In November 2024 we participated in the [blank for review] (LUL) festival. This is a city-wide art trail which attracted over 80,000 visitors across three evenings. The festival's Creative Producer [blank for review] commented "Shadowplay fascinated and enthralled audiences, capturing their imaginations and intriguing them at the same time". The LUL installation incorporated a control desk position (Fig 2.2) and was our first opportunity to use PromptJ in a live environment with an audience. See a short video showcasing the installation at LUL here: [blank for review].

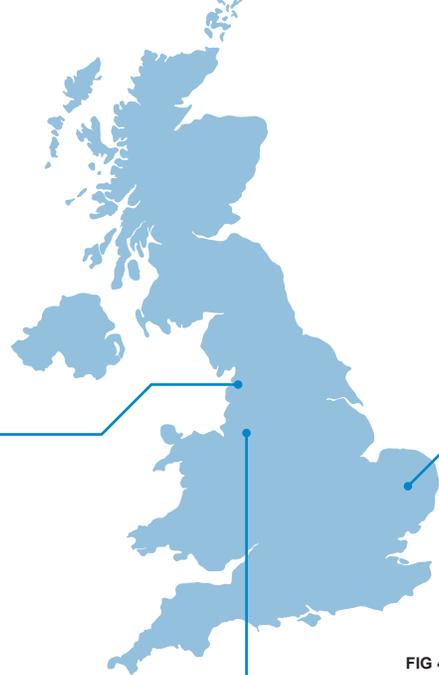

FIG 4.4

**SuperMassive Installation**

In September 2024 we installed Shadowplay at SuperMassive's inaugural exhibition in a disused unit in a public shopping centre in Salford, UK. SuperMassive are a startup company who build environments that leverage sound and light to create unique experiences. Ours was one of several installations that made up a sci-fi and space themed experience. This was our first experience of using PromptJ in the wild. In this case we used PromptJ to fine-tune and tweak the Shadowplay experience aligning it to the aesthetic of the rest of the exhibition, however PromptJ was then left in an autonomous 'autopilot' mode.

**KlangHaus collaboration**

In December 2024 we collaborated with KlangHaus. KlangHaus are an art rock group comprising a band and a filmmaker. We visited their rehearsal studio in Norwich (which is also a performance venue) to demonstrate and experiment with Shadowplay. The intention was to explore if and how KlangHaus could use PromptJ to incorporate Shadowplay into their performances. We also wanted to learn from witnessing how a new user would interact with PromptJ. In this case that user was KlangHaus's filmmaker Sal Pittman (see https://salpittman.com/). KlangHaus used PromptJ and Shadowplay to produce a significant amount of imagery (e.g., Fig 4.3, Fig 7.2) which was then recorded as videos and utilised in a series of live performances.

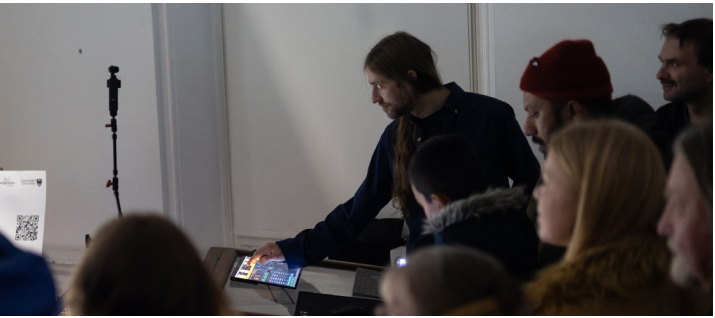

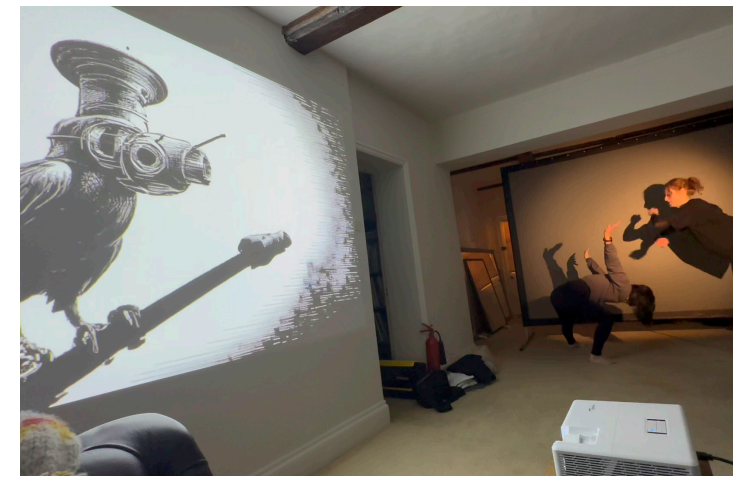

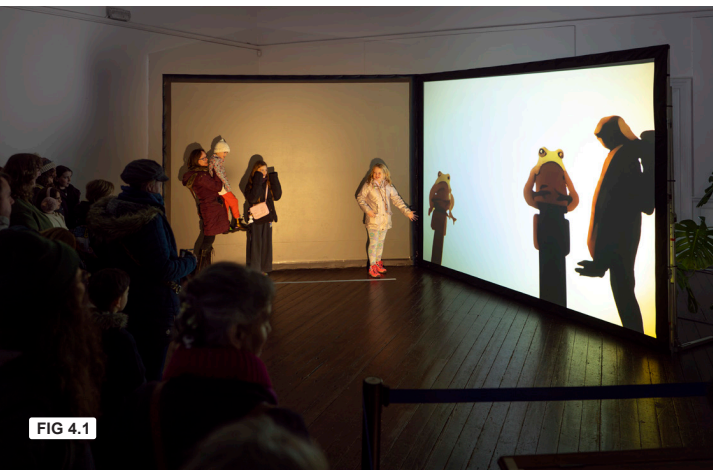

FIG 4.1

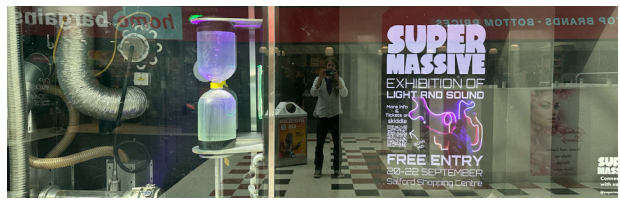

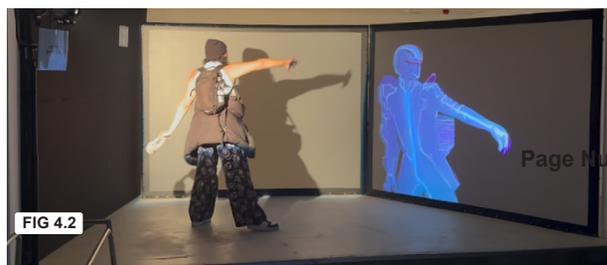

FIG 4.2

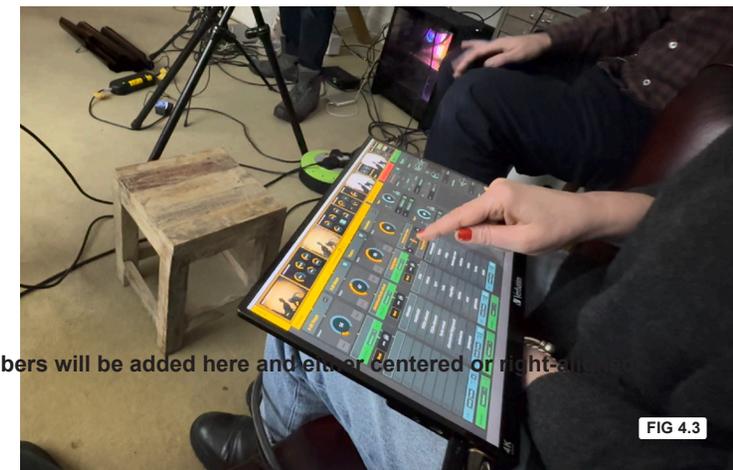

FIG 4.3

## REFLECTIONS, LEARNINGS & STRONG CONCEPTS

The subsequent pages discuss a series of strong concepts [8]. Our interpretation of strong concepts, also referred to as intermediate knowledge, is a type of knowledge that sits on a continuum between insights relating to highly specific design instances and generalisable and universal theories. Design Research processes frequently create this kind of knowledge and it is characterised as ideas that, while they are inferred from specific examples, can be applied to other contexts. Considered together, the following strong concepts constitute the first of two contributions this pictorial makes. They are presented in a way that is meant to prioritise ease of comprehension (rather than chronologically, or according to a thematic categorisation). The pictorial's second contribution–towards the end of the pictorial–is a novel prototype design which builds upon these strong concepts.

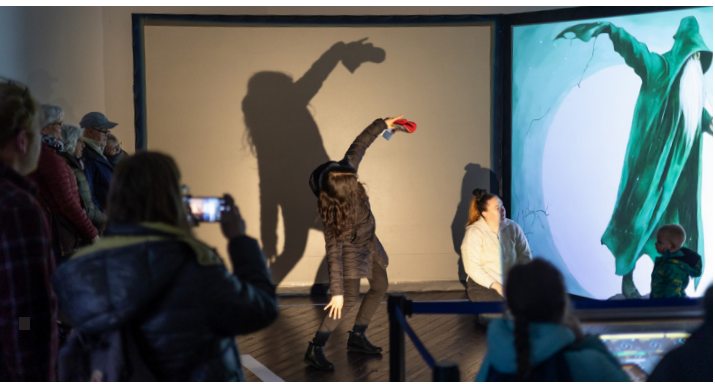

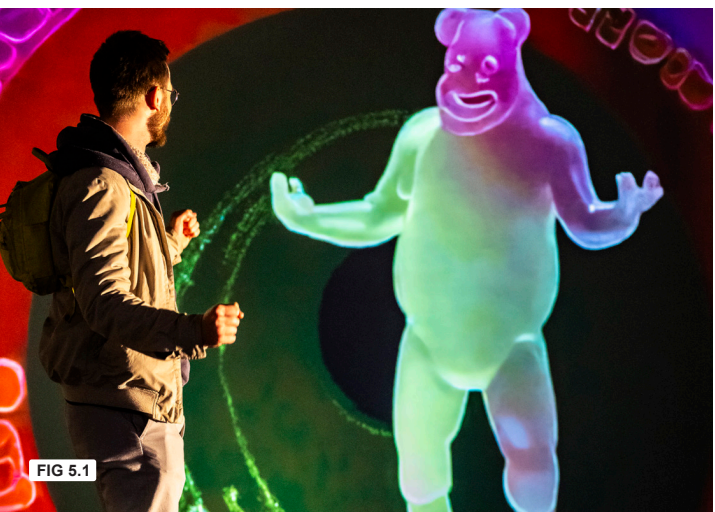

FIG 5.1

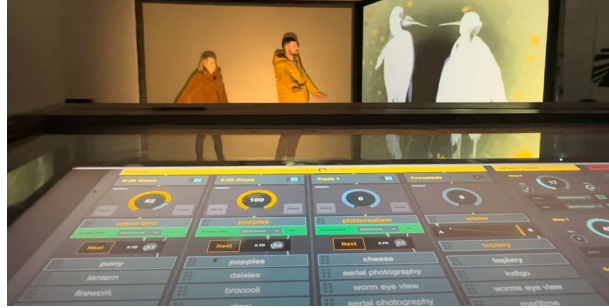

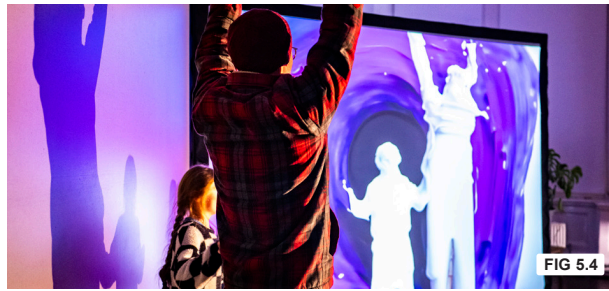

FIG 5.4

### Performance and creative exploration in the same UI

The PromptJ interface proved itself to be useful for performance and creative exploration. At LUL we used PromptJ to deliver live performances, responding to the audience, authoring text prompts and manipulating parameters live. This allowed unique imagery to be created responsively on an ad hoc basis. Meanwhile, for the SuperMassive installation we used precisely the same UI to rapidly curate a bespoke settings and text prompts that, together, created a version of Shadowplay that was sympathetic to the aesthetic of the overall exhibition but ran in an autopilot mode without any need for intervention. Our work with KlangHaus combined elements of performance and creative exploration, with mini-performances testing newly-developed combinations of settings and text prompts, which were then fine-tuned. In sum, the mixer skeuomorph as a means to interact with GenAI seems to be an intuitive and flexible heuristic that can allow both performance and creative exploration through the same UI.

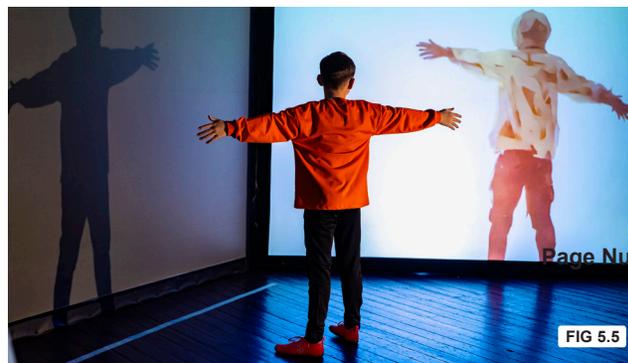

FIG 5.5

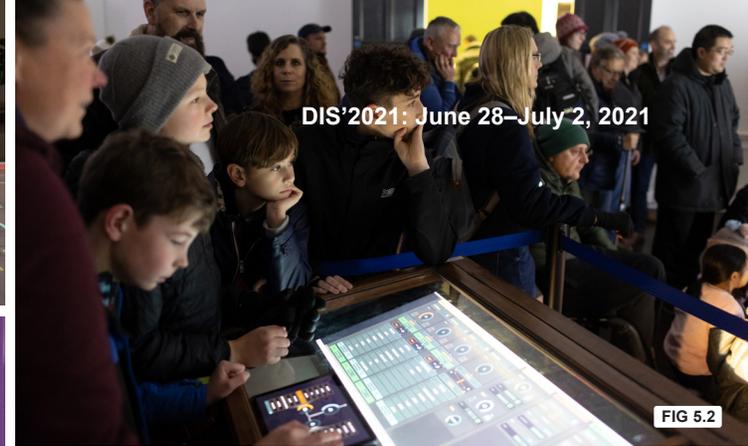

FIG 5.2

### Simple controls can achieve powerful results

At the LUL installation, the PromptJ control desk was a prominent feature. On the first of the 3 days, we invited visitors to try out the UI (Fig 3.1). While visitors enjoyed having control, it was clear that PromptJ's abundance of dials and sliders, each relating to different aspects of the image-production, was challenging for most novice users. The usability challenge was not just the amount of controls, but that it is hard to visualise the relations between them. To address this we rapidly created a simplified UI prototype that could be used via a wireless touchscreen for the subsequent days. It allowed visitors access to just the basic weight control of two prompt channels. It became evident that while a small fraction of the overall control parameters wfeere available, visitors could still access a gratifying variety of creative space. One particularly keen young visitor controlled the installation for around 60 uninterrupted minutes, deftly co-producing (with those casting shadows) a varied and engaging performance.

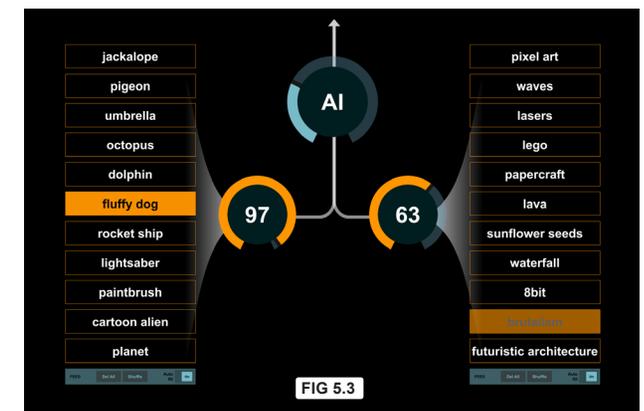

FIG 5.3

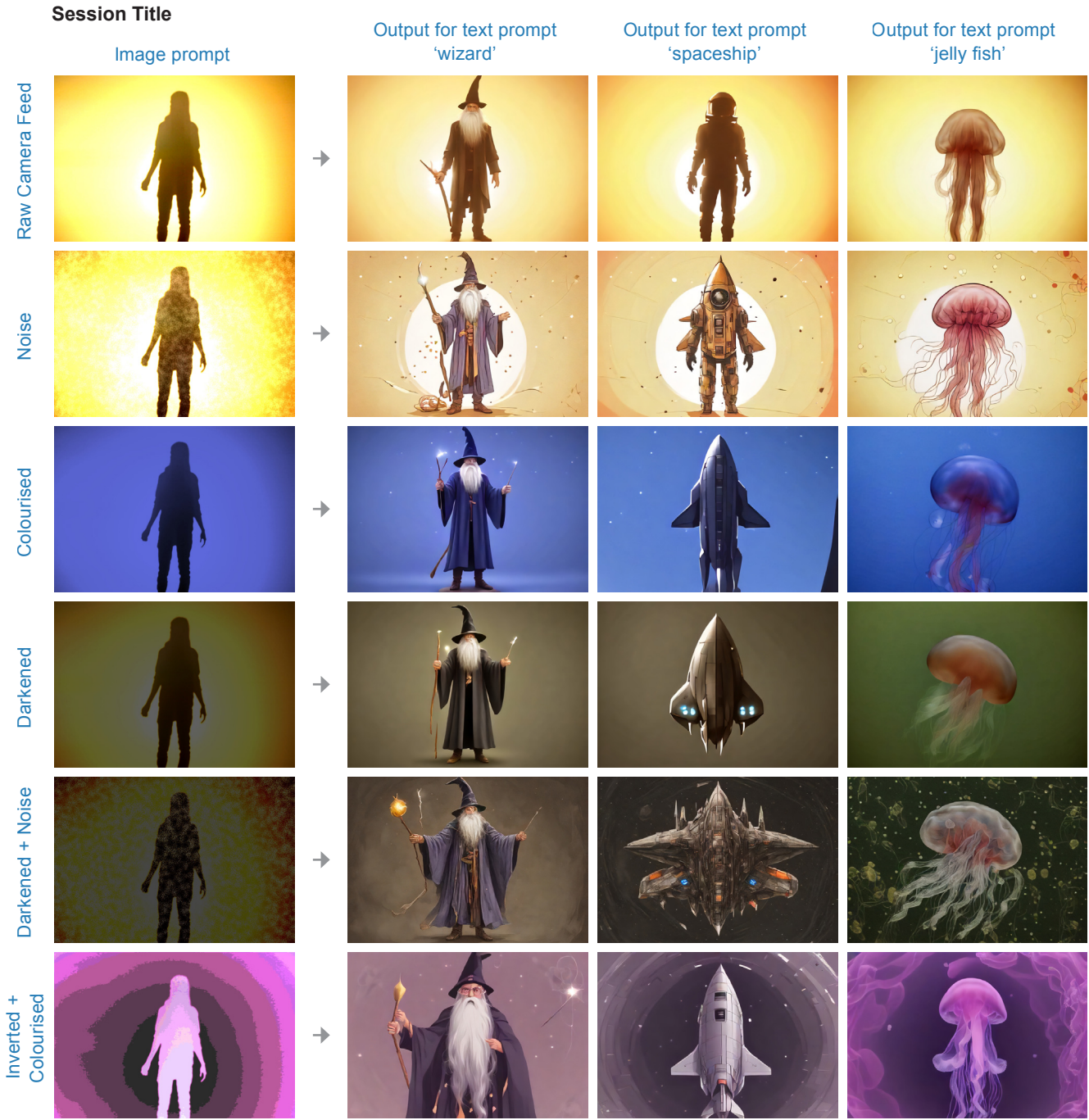

FIG 6.1

**Not all text prompts are equal**

The skeuomorph mixer design of PromptJ (Fig 3.1) quickly revealed that not all prompts are equal; some prompts are much stronger than others. For example, the text prompt 'frog' is highly prominent in the model. Therefore, if the mixer was used to blend frog with a less prominent prompt, e.g., 'sculpture', on another mixer channel, then it would be clear that equal weight settings did not produced a visually balanced output. Hence, PromptJ was very good at highlighting imbalances within the model, and once highlighted, facilitating easy compensation as desired by the user.

**Don't underestimate the power of pixel prompting**

Light prompting is the use of "light and shadows to create input images whose attributes provide increased control over the latent space in the model" [1]. In the case of Shadowplay, this is achieved by creating shadows, capturing those with a camera, and then using the resulting image as the input. However, we note that from the model's perspective, the notion that it is light driving this effect is irrelevant. The model is simply processing input pixels, regardless of their source. Hence, perhaps pixel prompting is a better term.

In PromptJ we added controls to alter brightness, contrast, colour balance and noise level of the input image. While we added these controls because we knew they would extend creative possibilities, when collaborating with KlangHaus we recognised a much greater potential for these controls to be used in shaping the output image. Arguably, pixel prompting, i.e., adjusting properties of the image like brightness or noise levels, is of equal importance to the actual content of the input image. Pixel prompting was also in play with SuperMassive, to match their futuristic/space theme, we created a bespoke mode that manipulated pixels by inverting the image, increasing the contrast, and forcing artificial colours into the bright areas (which, due to the inversion were the shadows captured by the camera). The result was high-contrast, character-focused imagery with a consistent aesthetic.

## Anchors for consistency

PromptJ was intentionally designed as a tool that allows a user to navigate through the gamut of the model's latent space. In production contexts, however, endless possibilities are not always desirable and some amount of consistency and control is required. It is useful to consider across all the parameters that affect the output image, which are the parameters (or combinations thereof) that 'anchor' the output in a consistent manner. Such anchors can be text prompts, pixel prompts, content of the input image or assemblies of all of these.

During our collaboration with KlangHaus a circular motif became a consistent element in the outputs. This originally came from a bright spot in the lighting used to make Shadowplay work in this environment, however, that became an inspiration for a series of text prompts. Those text prompts then reinforced the circularity and contrast of the bright spot. Finally, further pixel prompt tweaking, in particular the introduction of high-frequency/low-gain noise to the image, added more aesthetic stability. Together these factors acted as an anchor for consistency.

In the SuperMassive installation, we were striving to create a vivid, colourful, character-driven output image. To achieve this our anchors were based on the combination of settings in Salford Mode (see Don't underestimate the power of pixel prompting, page 6). Our workflow involved establishing these pixel prompting anchors, then testing a wide array of text prompts and keeping those which produced desired results, hence we selected our text prompts to be compatible with the pixel prompt anchors.

Another way of considering anchors relates to strong signals in the original input image. For example, we noted how creating horizontal shadows would, almost without exception, create the effect of a horizon in the output image. This effect was significant, on parity with text prompting or other types of pixel prompting.

There are likely numerous ways to create consistency anchors, these examples are intended to demonstrate the concept and suggest starting points to experiment from.

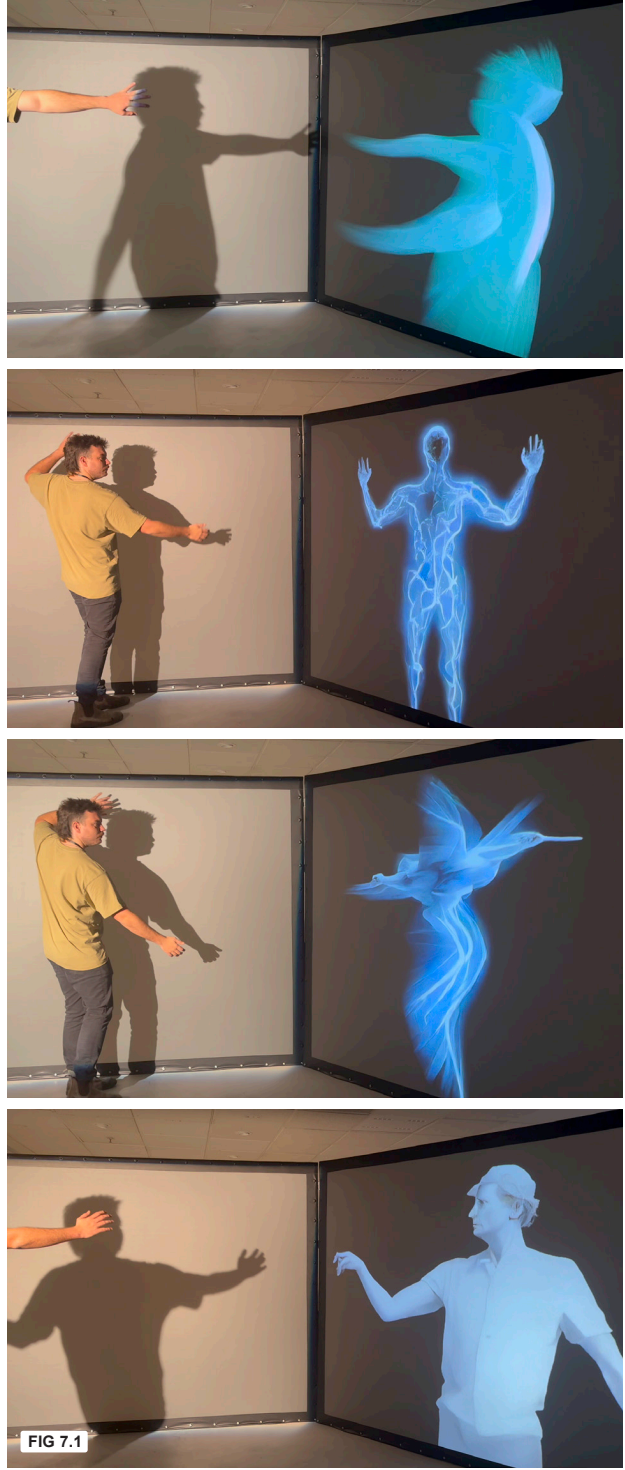

FIG 7.1

IMAGE PROMPT    MODEL OUTPUT

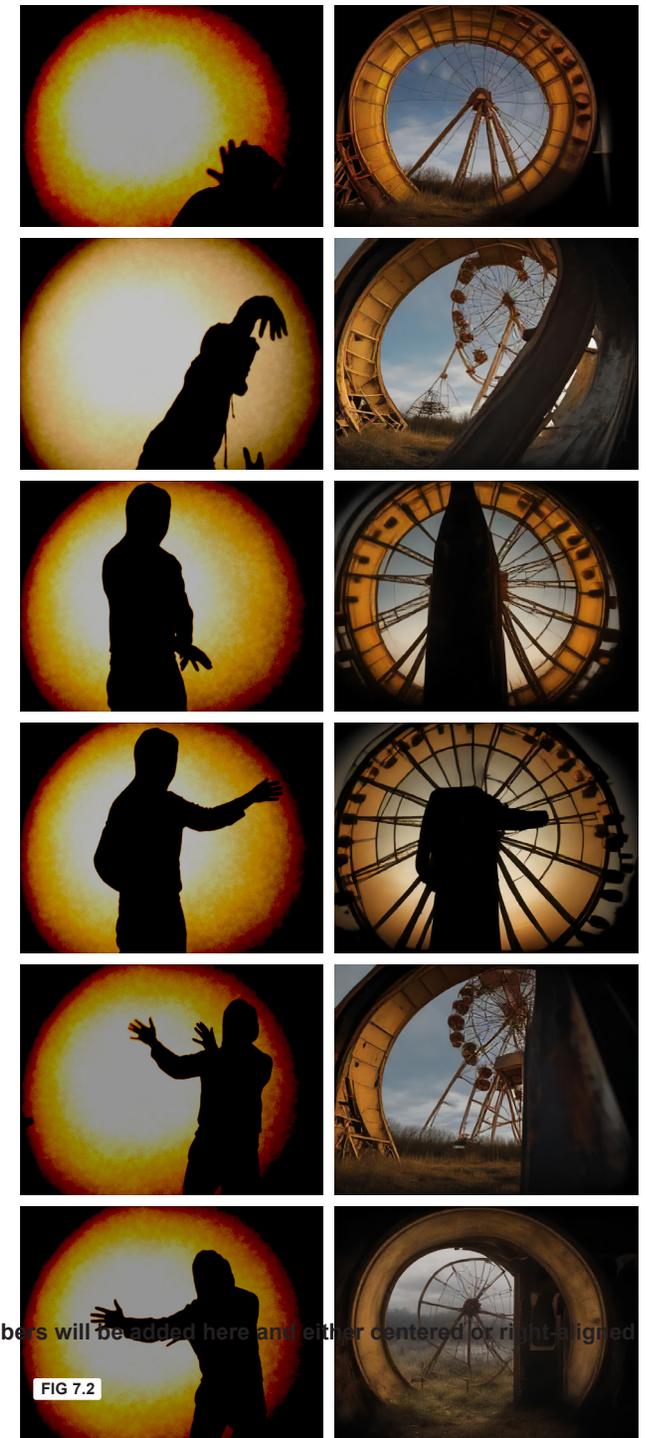

FIG 7.2

## Reducing cognitive load through automation

PromptJ offers almost limitless possibilities for blending, balancing, and exploring text and pixel prompts. While these can all be operated completely manually, we anticipated the need for, and implemented, some ways to automate aspects of interactions. The most obvious example is the crossfade feature. Within PromptJ, on each channel a user can define a crossfade ('X FD', see Fig 3.1) time. This is the number of seconds for a fade to complete between an existing and a new prompt on any given channel. The same could be accomplished by employing multiple channels, just as it can be achieved in an audio mixer by using two volume controls, however, by automating this feature cognitive loading is reduced which facilitates a focus on expression rather than micro-managing controls.

We conducted several experiments around similar automations of other parameters in the workflow with a range of triggers or modulators. For example, adding a low-frequency oscillators to modulate colour controls and noise parameters and, with KlangHaus, connecting the amplitude of an incoming audio signal (in this case coming from live piano) to the brightness parameter.

In a completely different type of automation we implemented a pluraliser that detected the number of people in a given input image and amended the text prompt accordingly. For example, if the underlying text prompt was 'priests' but two people were detected in the input image, then the text prompt would be amended to 'two priests' (Fig 8.4). In our tests this strongly reinforced the definition of characters in the output images, which was a useful effect for our purposes, but highlighted how any input (e.g., detecting the colour-tone of the input image) could be used to drive the text prompt (e.g., if the input is blue, add underwater or sad to the prompt).

While the rationale for PromptJ was manual control, it was clear that in some contexts automation would actually enhance and enable expressivity.

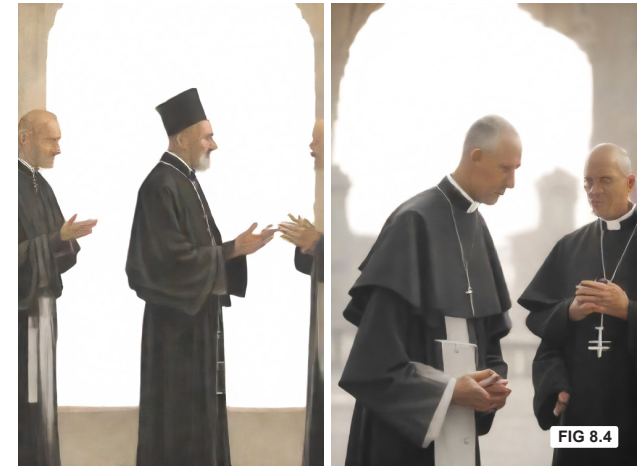

Text prompt 'priests'    Text prompt 'two priests'

FIG 8.4

Below, three automated sequences produced through automation varying brightness (Fig 8.1), diffusion level (Fig 8.2) and a automated prompt cross fade ('woolen yarn' to 'folded paper', Fig 8.3)

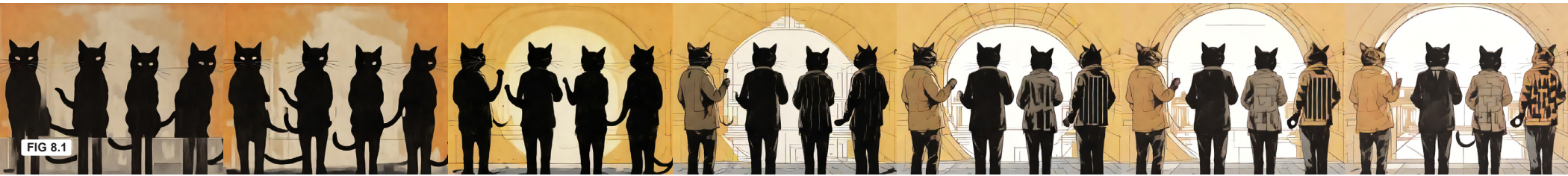

FIG 8.1

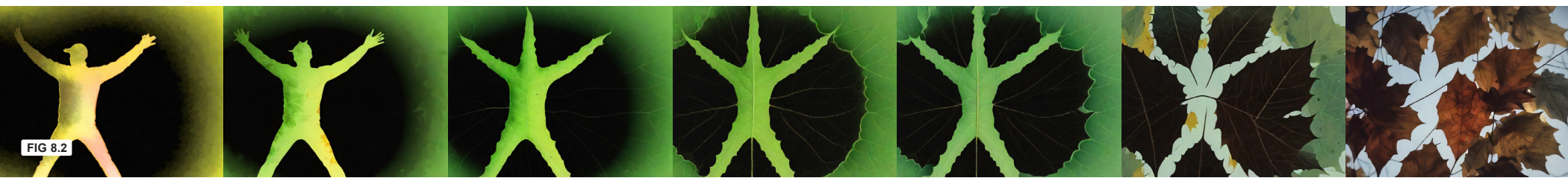

FIG 8.2

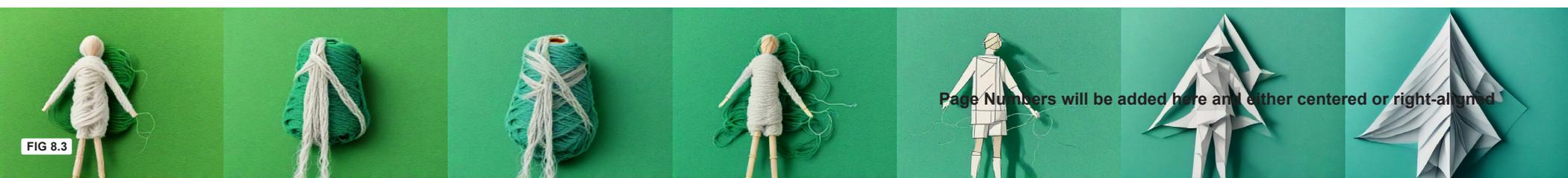

FIG 8.3

**HOLISTIC PROMPT CRAFT**

The rapid speed of adoption and development in this area necessitates a methodology capable of a rapid response, arguably Research through Design or practice-based Design Research meets this need (see [3]). The strong concepts emerging from our work are not intended to be interpreted as generalisable theories, recommendations or even implications for design. Instead, as intermediate knowledge, they should be interpreted as inspirational or generative ideas that are primarily relevant to others who are designing interactions that involve GenAI. It is at this confluence of contemporary practice and proposal for future inquiry that this work's timeliness and significance comes into focus. Within the context of the pictorial this page serves to two roles: (1) contextualise the strong concepts and describe an abstract connection between them into a single framing which we call Holistic Prompt Craft; and (2) provide a bridge to the pictorial's second contribution, a prototype UI that embraces the Holistic Prompt Craft framing. We call this UI PromptTank.

While the relationship between input image and outputs in image-to-image GenAI pipelines may superficially appear straightforward (text prompts and input images influence the output), the strong concepts we have presented describe a range of less obvious interaction affordances and opportunities that this new class of technology imports. Current approaches to GenAI interfaces typically fall into two extremes. Most commonly, systems offer little more than a text input box and a few predefined parameters, leaving users to explore the model's capabilities through trial and error with text prompts alone. At the other end of the spectrum, advanced interfaces employ a 'wires and boxes' paradigm (e.g., ComfyUI), allowing users to construct complex custom pipelines. While this latter approach offers granular control, its complexity makes it inaccessible to non-expert users.

As demonstrated by our use of PromptJ, there are other ways to interact. Something as simple as range-slider or fader control to manipulate the weighting of text prompts is a viable and useful way of making an otherwise unremarkable text prompting process highly interactive. Through that same interaction a creative potential is opened up, a potential that is useful for both crafting of text prompts as well as live performances. Such interactions expose the internal nuances of models in a way that is more 'tactile' than text box-only or wires-and-boxes interfaces.

Another equally important aspect of this, particularly relevant to image-to-image GenAI systems, is the notion of pixel prompting. While the broad influence of input images on outputs in image-to-image systems is straightforward, our work with PromptJ and Shadowplay uncovers subtle yet powerful mechanisms. Elements of input images that human perception tends to normalize, e.g., such as noise patterns, contrast ratios, or tiny variations in brightness, can dramatically shape the output. This is entirely new to GenAI, and hence it is important to help support users in recognising it, learning how to harness it, and utilising it.

Previously we framed Prompt Craft as "a craft-like navigation of the latent space within generative AI models" [1]. However, that notoin was constrained to the development of text prompts. PromptJ's mixer-inspired skeuomorphic design concept highlights, however, that the craft like framing can be expanded to include not just text prompting but everything that impacts the output images produced during an interaction. This holistic view, which we term *Holistic Prompt Craft*, is an integrated stance. Hence, it is not sufficient just to be aware that both text prompts and pixel prompts are relevant, but rather to embrace that they are entwined and interlinked with one another, and with the full range of elements that impact the output images, including the content of the input images and affordances of the underlying model. This small conceptual shift allows UI designs that can support users in understanding and using the significant parameters, and in turn to access creative juxtapositions and supportive reinforcements within models' latent space that might otherwise remain inaccessible.

While it makes the complexity of the technology conceptually available, Holistic Prompt Craft does not necessitate complexity in use. Simple controls can, after all, be enough for a large amount of expression (see Simple controls can achieve powerful results, page 5) meanwhile many significant parameters are likely to become fixed in any given use case and become anchors (see Anchors for consistency, page 7). The strength of this approach lies not in simultaneously manipulating all available parameters, but in understanding their relationships, potential, and making them available for users to exploit.

To more explicitly and visually demonstrate what we mean by Holistic Prompt Craft, further developing the notion, the pictorial's second contribution is the PromptTank prototype UI design. PromptTank is currently a functional alpha, but has not been tested in the wild. While PromptJ performed very well across the contexts it has been used, and has been an excellent vehicle for developing the strong concepts which cohere as Holistic Prompt Craft, the way we implemented the mixer concept has various limitations. For this reason PromptTank reimagines how a user might prompt and interact with GenAI models in several key ways. These include: a spatial approach to how a user may ideate and organise prompts prior to use; a single weighting/interaction mechanic of parameters supporting multiple interactions or users at a time; connection of prompts together to form nodes that can influence text and pixel prompting at the same time; easy-to-implement automation embedded within nodes.

An annotated wireframe of PromptTank appears on the following page and a discussion of how it reflects Holistic Prompt Craft is further unpacked in the pictorial discussion section (page 11 and 12).

**THE PROMPTTANK UI**

*Prompt nodes*
The basic metaphor of the PromptTank UI is of prompt nodes – circular objects that are moved around to form a weighted prompt

*Image prompt controls*
A range of adjustment as offered in a palette of adjustment nodes, and once selected appear in the active zone. We have prioritised the most useful adjustments, and made each adjustable along the common weighting axis for all nodes.

*Autopilot controls*
Autopilot mode is presented minimally, with an on-off switch along with controls for frequency of prompt changes and crossfade times.

*Model controls*
Key parameters for controlling the model are presented for easy thumb control on a tablet.

*Active zone*
Nodes moved to the upper 1/3 of the UI are made active. The nearer to the top, the higher the weighting.

*Storage zone*
Nodes in the lower 2/3 of the UI have zero weighting, and can be organised spatially by the user.

The spatial balance between the active and storage zone can be adjusted, but this 1/3 to 2/3 setting gives plenty of space for ideating and organising prompts.

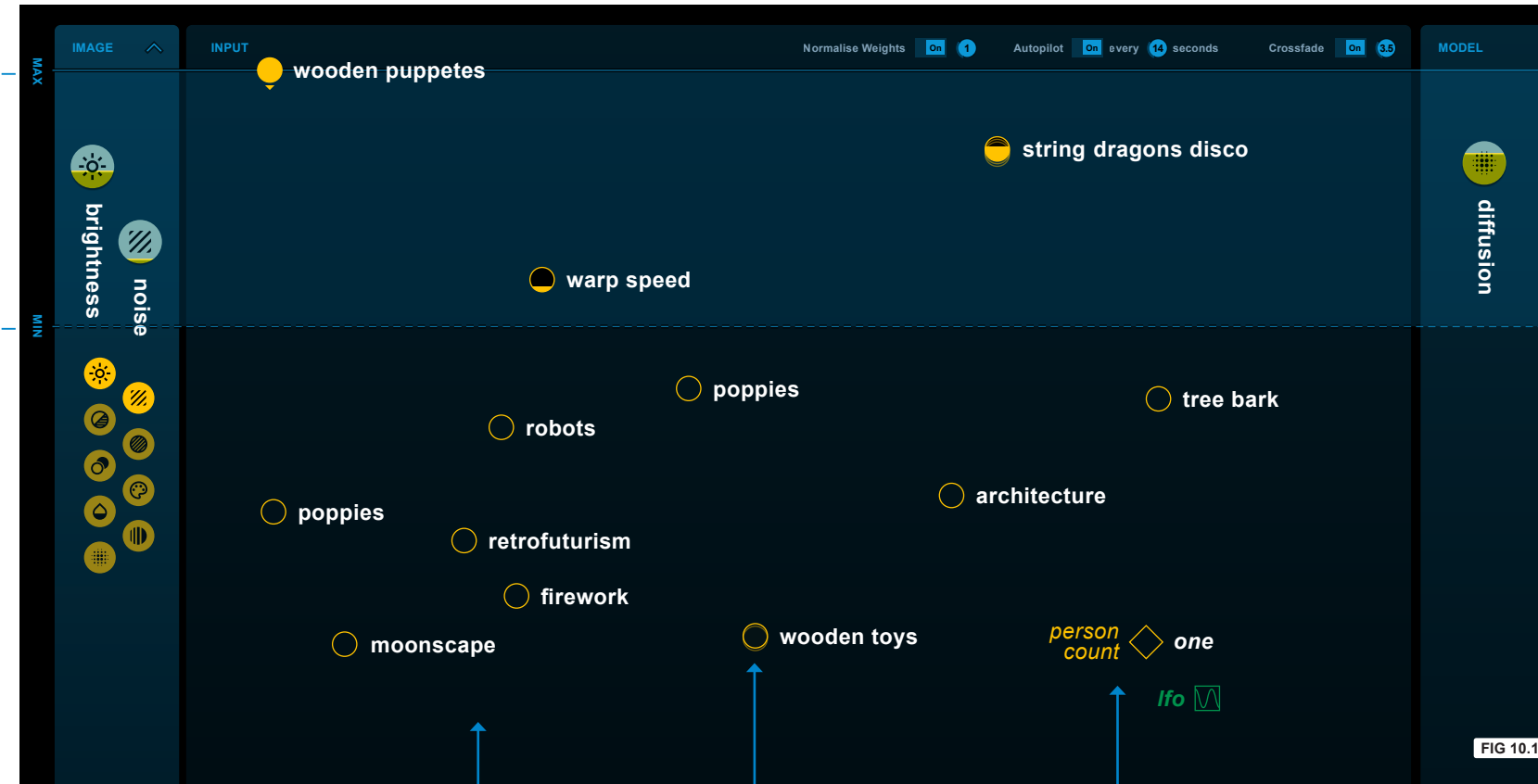

FIG 10.1

*Continuous text entry*
Keyboard or voice entry of prompts is constantly active, allowing for new prompts to be added to the storage zone at any moment.

*Text prompt nodes*
The majority of the UI is given over to a open space for text prompt nodes that can be freely repositioned through drag gestures.

*Joining nodes*
Multitouch drag gestures allow nodes from any part of the UI to be combined, facilitating rapid experimentation with text prompts

*Automated nodes*
Specialised nodes can be joined to others to produce automation effects, such as a text node that reflects the number of human figures detected in the input image.

**Page Numbers will be added here and either centered or right-aligned**

## DISCUSSION

VIa our practice-based Design Research approach, the empirical basis for this work is our experience of designing and using the PromptJ interface (Fig 3.1) to enable several instances of the Shadowplay installation. Integral to Shadowplay is a GenAI image-to-image stack, which generates frames at around 12 frames per second. The speed of generation gives the effect of a video, hence terming it 'real-time'. Despite this effect, at generation time each frame goes through an end-to-end process based on the current set of parameters and there is no technical linkage or persistence between frames. Hence this is technically no different to any other diffusion-based AI image generator. The strong concepts that we presented in this pictorial are, therefore, directly relevant to others designing with or for real-time image-generation pipelines but we posit our strong concepts and the notion of Holistic Prompt Craft are likely to be relevant to GenAI much more broadly.

As UIs for image-focused GenAI matures, it is plausible (or perhaps likely) that real-time feedback showing how the manipulation of parameters will affect the output will become a ubiquitously available feature. Even if the end goal is the generation of individual static images, as is the case in many contemporary systems (e.g., MidJourney, Adobe Firefly, etc), real-time feedback on the effect of parameters in a preview window will become within reach on a technical level as compute and models become more efficient. In such a scenario, the significance of our contributions expands beyond systems that are designed exclusively for real-time applications to include any image-focused GenAI system.

Similarly, as GenAI models move towards multi-modality our findings may be relevant outside of the realm of image generation. The commonplace chat interfaces that define the current state of the art in GenAI act to almost entirely obfuscate the complexity of the underlying model or models that produce the output. As applications of GenAI mature finer-grained control over underlying parameters (e.g., from temperature or randomness controls through to adherence to alignment and safeguard settings) may need to be exposed. Enabled by increasingly fast inference times, real-time tweaking of such parameters would mean that the logic underpinning the notion of Holistic Prompt Craft would be relevant in this domain also.

Whilst these scenarios are speculative, they demonstrate that this work is significant and not only applies to the current state of the art in real-time GenAI image generation, but to the potential future of AI image generators, and interaction with GenAI more broadly. Moreover, these suggestions align with the inspirational and generative intent behind our contributions' framing as generative and/or inspirational strong concepts.

### PromptJ to PromptTank via Holistic Prompt Craft

When cohering our strong concepts into the umbrella notion of Holistic Prompt Craft (see page 9) we explained how the PromptTank design concept was created with both our strong concepts and broader idea of Holistic Prompt Craft in mind. Here we provide some more detailed reflections on the intersection between what we learned through using PromptJ (the ideas collated into the strong concepts), how those ideas are manifested in the PromptTank concept, and what this reveals about the Holistic Prompt Craft framing.

One significant limitation of PromptJ was how it handled ideation and queuing of text prompts. Due to the constraint of our implementation of the mixer skeuomorph, only 9 text prompts were concurrently visible for any given channel. Usually each channel would have many more prompts loaded into it, but they would not be visible at the same time and could only be accessed by pressing 'Next' (see Fig 3.1). Through our own and others' prompt-creation sessions, we noted how it was often useful to use channels to group prompts according to their semantic purpose. For example, one channel could be used to house prompts that were used as styles such as painting, glass, or papercraft while another channel could contain subjects such as elephant, planet, or priest. While this categorisation was useful, invariably circumstances would arise where the categories needed to be mixed and amended on the fly. To achieve this prompts would have to be copied into a third channel such that they could be mixed appropriately. Hence, while the mixer metaphor was exceptional at balancing prompt weights it was less good at supporting this kind of conceptual flexibility with how prompts can be used in multiple ways and in a plethora of different combinations. PromptTank addresses this by allowing text prompts to be organised between the storage zone and active zone in a fluid and flexible manner. While addressing a limitation of PromptJ, this simple tweak is also indicative of how Holistic Prompt Craft offers a novel perspective. Given that text prompts can play a wide array of different functional roles in how any given image is created, it is important that a user has flexibility of how to organise them, envisage their role in the overall image generation process, and conceptualise their relationships to other parameters.

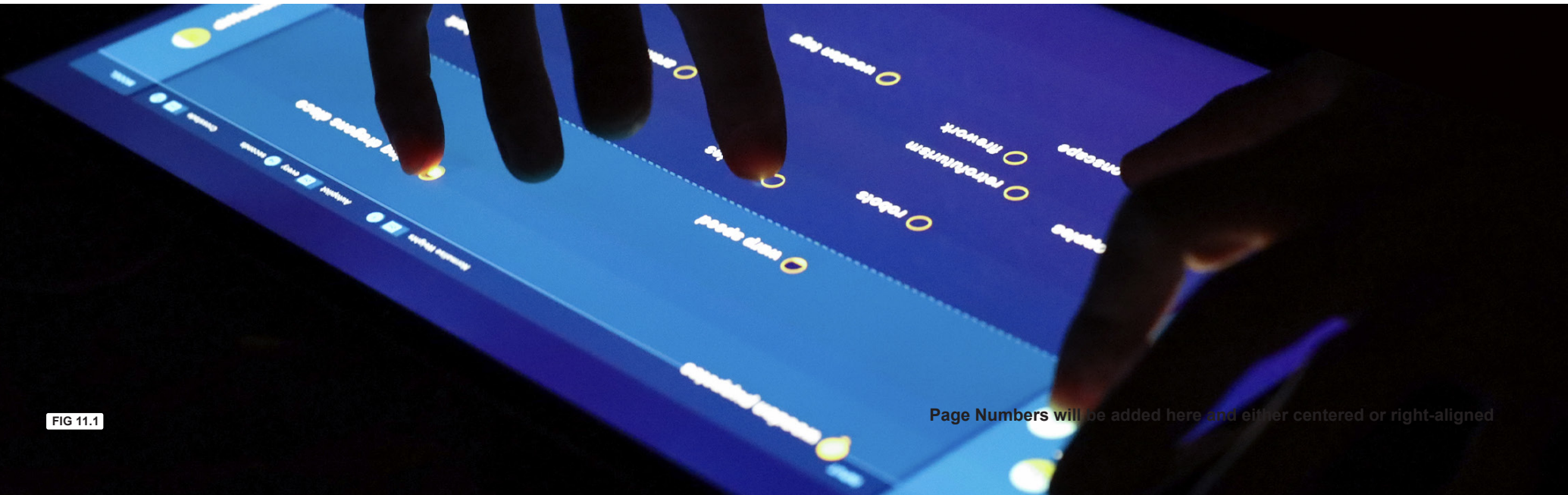

FIG 11.1

Expanding upon the reality of Holistic Prompt Craft further, it is pertinent to remember this does not simply mean be aware of both pixel prompting and text prompting, but to appreciate that every significant parameter (e.g., the source image, image processing, text prompts/weights, diffusion steps, etc) could have an impactful relationship with every other parameter. Because the exact way these relationships play out is a product of the complexity within the AI model, the relationships are not deterministic or predictable. For example, while increasing the brightness in Fig 8.1 made the feline figures have clothing, this would not have been the case with another input image or parameter setting. This non-deterministic property is, in our view, one of the key reasons why a more unified or holistic mindset is useful in prompt craft. These ideas were represented in the PromptTank prototype through a less linear and hierarchical way of organising parameters. While PromptJ communicates how the data flows through the system in a linear manner that is true to how the underlying technology works (see Fig 2.1, 3.1), with PromptTank we dispensed with this in favour of allowing the user to place any node wherever they like regardless of which part of the signal path it relates to. This means semantically similar nodes can be placed next to each other and reorganised on the fly, while avoiding the unnecessary constraints of the mixer metaphor while preserving all the same functionality.

PromptTank is currently in a functional alpha stage and is having positive early feedback regarding dynamic prompt grouping, manipulating multiple active prompts (using multi-touch) and leveraging the full gamut of parameters. The prototype suggests that Holistic Prompt Craft is a useful way to represent the amalgamation of the strong concepts we described based on using PromptJ, but also can serve as a conceptual jumping off point or inspiration for the design of UIs for interacting with GenAI. Future work will evaluate PromptTank more fully, explore different use cases (e.g., performance, prompt development, education) of PromptTank, and extend features (e.g., turning the active area to be an infinitely-scrollable canvas). We have also discussed and sketched the idea of 'nested tanks'. In this scenario the given state of any PromptTank instance would be saved as a preset, which would then be available as a node in its own right. This would then allow a user to easily blend, crossfade, and manage a multitude of complex creative possibilities.

**Timeliness and significance**

The rate at which GenAI has been adopted is unprecedented, and this is currently represented in the prevalence of research that either uses or is about GenAI at DIS and the broader SIGCHI family of conferences. Framing the relative significance and timeliness of this work, we suggest that the current era of GenAI UIs are likely to be the most rudimentary and unsophisticated versions that we will see, compared to yet-to-be-developed interfaces of the future. The strong concepts and prototype design we present here are not intended as didactic. Rather, our contributions are intended to channel the experiences and insights gained through our deployments of PromptJ and Shadowplay into practical know-how (our strong concepts), a higher level abstraction for ways of thinking about interaction with GenAI (Holistic Prompt Craft), along with specific design concepts (PromptJ and PromptTank).

**SUMMARY**

This practice-based Design Research project explores the affordances of interactions with real-time image generators based on three installations of the Shadowplay system that utilised the mixer-inspired PromptJ UI. Our first contribution reflected on these installations to produce a series of strong concepts for designing with real-time image generation. We then cohered and framed those concepts with the Holistic Prompt Craft metaphor. Our second contribution exemplifies and embodies Holistic Prompt Craft in the PromptTank UI prototype. Our intention in sharing this work is to contribute to the rapid development of GenAI applications and interfaces designed to support them.